
\magnification=\magstep1
\hsize=384pt
\vsize=580pt
\lineskip=1.5pt
\lineskiplimit=1.5pt
\parskip=3pt plus 1.5 pt
\parindent=1pc
\def\zhs{\hskip 0pt plus 0pt}
\def\.{\vphantom{x}.}
\def\"#1{{\accent"7F #1\penalty10000\zhs}}
\font\gigant=cmr10 scaled \magstep3 
\font\gross=cmr12 
\font\mittel=cmbx10 scaled \magstep1 
\font\fett=cmbx10
\font\eightrm=cmr7 
\font\name=cmcsc10
\def\rmfolio{{\rm\folio}}

\headline={\ifnum\pageno=\firstpage\hss
           \else\vbox
             {\line{\hbox{\vrule height 7.8pt depth 3.2pt width 0pt}
                    {\name\hfill page \rmfolio}
                   }
              \hrule
              \hrule height 0.6pt width 0pt
             }
           \hss\fi}

\newcount\fnn
\def\nospaceskip{\spaceskip=1sp plus 1sp minus 1sp}
\long\def\fusn#1{\global\advance\fnn by 1
                 {\nospaceskip
                  \baselineskip=9pt
                  \setbox\strutbox=\hbox{\vrule height 7pt depth 2pt width 0pt}
                  \eightrm
                  \everypar{\hangindent=\parindent}
                  \footnote{$^{\number\fnn}$}{#1}
                  \everypar{}
                 }
                }
\def\footnoterule{\kern-3pt
                  \hrule width \hsize
                  \kern2.6pt}
\let\sst=\scriptstyle
\newcount\firstpage
\newcount\kn
\newcount\fn
\edef\kno{}
\def\mkkno{\xdef\kno{\number\kn}}
\def\kapinit#1#2{\global\kn=#1\global\pageno=#2
                   \global\fn=0
                   \mkkno}
\def\nextkap{\global\advance\kn by 1
             \mkkno}
\edef\kus{}
\def\kusout#1{\xdef\kus{\kno. #1}
         \medskip\vskip 2.7pc\leftline{\mittel\kus}
              \medskip} 
\def\aus#1{\bigbreak\leftline{\fett #1}
            \nobreak\smallskip} 
\def\gln{\number\fn}
\def\rgln{{\rm\gln}}
\def\glasno#1{\global\advance\fn by 1
              \expandafter\xdef\csname#1\endcsname{\gln}
              }
\def\gl#1#2{\glasno{#1}$$#2\eqno{(\rgln)}$$}
\def\glalign#1#2{{\glasno{#1}
                  \def\ncr##1{&{\rm (##1)}\endline}
                  \def\cr{&\endline}
                  $$\eqalignno{#2}$$}}
%
%
\def\kasten#1#2#3{\mathord{\mkern0.5\thinmuskip
                           \vbox{\hrule height 0.3pt
                                 \hbox{\vrule width #3
                                       \hskip#1
                                       \vrule height #2 width 0pt
                                       \vrule width #3}
                                 \hrule height 0.3pt}
                           \mkern0.5\thinmuskip}}

\def\intd#1{\int {\rm d}^4#1\,}

\def\mdef{\mathrel{\raise 0.25pt\hbox{:}\!=}}
\def\mfed{\mathrel{=\!\raise 0.25pt\hbox{:}}}
\def\al{\alpha}

\def\ga{\gamma}
\def\de{\delta}
\def\pd{\partial}
\def\ep{\epsilon}
\def\epv{\varepsilon}
\def\ze{\zeta}

\def\thv{\vartheta}

\def\ka{\kappa}
\def\la{\lambda}
\def\my{\mu}
\def\ny{\nu}
\def\rh{\rho}
\def\rhv{\varrho}
\def\si{\sigma}
\def\ta{\tau}
\def\yp{\upsilon}
\def\ph{\phi}
\def\phv{\varphi}
\def\ch{\chi}
\def\ps{\psi}
\def\om{\omega}
\def\halb{{\textstyle {1\over 2}}}

\def\viertel{{\textstyle {1\over 4}}}

\def\d'al{\mathchoice{\kasten{4.9pt}{6.3pt}{1pt}}
                     {\kasten{4.9pt}{6.3pt}{1pt}}
                     {\kasten{3.4pt}{4.2pt}{0.7pt}}
                     {\kasten{2.4pt}{2.8pt}{0.5pt}}}
\def\T{\mathord{{\rm T}\,}}

\def\s{\mathord{{\sl s\,}}}
\def\Sl{{\cal S}\,}
\def\sC{\mathord{\,{\sl C}\,}}
\def\sT{\mathord{\,{\sl T}\,}}
\def\sCP{\mathord{\,{\sl CP}\,}}
\def\bra#1{\left<#1\right\vert}
\def\ket#1{\left\vert#1\right>}
\def\vev#1{\mathopen{\bra{0}}\mathinner{#1}\mathclose{\ket{0}}}
\def\vevt#1{\vev{\T #1}}
\def\grad#1{\left\vert#1\right\vert}
\def\reli#1{\mathord{\rlap{\raise
5.5pt\hbox{$\mathaccent"0224{\phantom{#1}}$}}#1}}

\def\pdrl{\,\reli\pd\,}

\def\sub#1{{\hbox{\eightrm #1}}}
\def\h{\hbar}
\def\Hi{{\it\Phi}}

\def\Ga{{\it\Gamma}}

\def\hi{\phv}
\def\go{\ch}
\def\cq{{\bar c}}
\def\fapo{{\rm gh}}
\def\psq{{\bar\psi}}
\def\ft#1{{\tilde{#1}}}

\def\rhq{{\bar\rh}}

\def\La{{\cal L}}
\def\Q{{\cal Q}}

\def\B{{\cal B}}
\def\Bn{\B}
\def\ord#1{{\cal O}(#1)}
\def\Ska{{\pd\cdotp\! A}}
\def\sla{\llap{/}}

\def\Asl{{A\sla}}
\def\ips{{i\epv}}
\def\pdsl{{\pd\sla}}
\def\pdrlsl{{\,\reli\pdsl\,}}

\def\Drlsl{{\,\reli{D\llap{/\hskip 0.167em}}\,}}

\def\LaFu{Lagrangian}
\def\Zimm{Zim\-mer\-mann}

\def\Lore{Lorentz}
\def\Loko{\Lore\ covariant}
\def\Abel{Abelian}
\def\FaPo{Fad\-deev-Po\-pov}
\def\HiKi{Higgs-Kib\-ble}
\def\Herm{Hermitian}
\def\FoRa{Fock space}
\def\SlId{Slavnov identity}
\def\GrFu#1{Green's function}

\def\Feyn{Feyn\-man}
\def\Geist{ghost} 

\def\WeZu{Wess-Zu\-mi\-no}
\def\Nonre{non-re\-nor\-mal\-iza\-bil\-ity}
\def\nonre{non-re\-nor\-mal\-iza\-ble}
\def\hien{high-en\-er\-gy}
\def\treun{tree-u\-ni\-tar}
\def\refI{[5]}
\def\refII{[7]}
\def\refIII{[9]}
\def\refIV{[8]}
\def\refVI{[6]}
\def\refJ{[1]}
\def\refGJ{[2]}
\def\refLA{[3]}
\def\refCLT{[10]}
\def\refLS{[11]}
\def\refDF{[12]}
\def\refWZ{[4]}
\def\abbI{fig.~I}   
\def\abbpaar{fig.~II} 
\def\abbII{fig.~III} 
\firstpage=1
\kapinit{0}{1}
\hrule height 0pt
\par\hfill\vbox{\hbox{SI 92-11}\hbox{}\hbox{October 1992}}
\vfill
\centerline{\gigant Cancellation of the}
\bigskip
\centerline{\gigant Chiral Anomaly}
\bigskip
{\noindent\gigant in a Model with Spontaneous Symmetry Breaking\vbox
      to 14.4pt{\hbox{{\bf 1}}\vss\vbox to 0pt{}}}
\par\noindent
\bigskip
\bigskip
\centerline{\gross Thomas PORTMANN}
\bigskip
\centerline{\gross Fachbereich Physik, Universit\"at-GH-Siegen,}
\medskip
\smallskip
\centerline{\gross Postfach 10 12 40, W-5900 Siegen, Germany}
\vfill
\aus{Abstract}
A perturbatively renormalized \Abel\ \HiKi\ model with a chirally coupled
fermion is considered. The \SlId\ is fulfilled to all orders of perturbation
theory, which is crucial for renormalizability in models with vector bosons.
BRS invariance, i.e.\ the validity of the identity, forces the chiral anomaly
to be cancelled by \WeZu\ counterterms. This procedure preserves the
renormalizability in the one-loop approximation but it violates the Froissart
bounds for partial wave amplitudes above some energy and destroys
renormalizability from the second order in $\h$ onwards due to the
counterterms.
\vfill
\hrule
\advance\fnn by 1
            {
                  \baselineskip=9pt
                  \setbox\strutbox=\hbox{\vrule height 7pt depth 2pt width 0pt}
                  \eightrm
\item{${}^1$}Work done at the Institut f\"ur Theoretische Physik,
                      Universit\"at Hannover, Appelstra\ss e~2,
                      W-3000~Hannover~1, Germany
                 }
\eject
\kusout{Introduction}
An approach to quantizing an {\it anomalous gauge theory\/} in the sense of
{\name Jackiw} \refJ\ is presented. The model under study is the simplest one
involving a chiral or $\ga_5$-anomaly in which all particles are massive, i.e.\
an \Abel\ \HiKi\ model with a chirally coupled Dirac fermion.

The spontaneous symmetry breaking (SSB) which is responsible for the masses
possesses three remarkable features. First of all, the mass of the vector boson
renders the model consistent although generally the anomaly violates the
conservation of the $U(1)$ charge beyond the tree approximation. However,
because of this one faces \Nonre\ due to the \hien\ behaviour of the massive
vector boson propagator. Instead, in this work it is suggested to write down
\WeZu\ counterterms by means of the inhomogeneous gauge transformation of the
Goldstone boson due to SSB and to cancel the anomaly by adding these terms to
the \LaFu. This preserves the gauge invariance of the model, keeps
renormalizability at least in the one-loop approximation, and therefore
provides a method of quantizing anomalous gauge theories.
Finally, the introduction of masses avoids problems related to infrared
singularities.

The same model has already been discussed by {\name Gross} and {\name Jackiw}
in section~III stage~ii of ref.\ \refGJ\ with the result that it is \nonre. In
addition, they considered the possibility of anomaly cancellation by those
counterterms and did not even then reach renormalizability. This paper confirms
that result to become relevant in the second order in perturbation. Those
linear combinations of counterterms are given which additionally become
necessary to define that order.

In contrast to ref.\ \refGJ\ which is based on the concepts and
techniques provided by {\name Lee} (for the model without fermion) and
{\name Adler} \refLA\ later approaches to the renormalization of gauge models
do not start any more from the gauge invariance of the regularization procedure
used. In the presence of an anomaly this is in any case not possible in a
consequent way. {\name Becchi}, {\name Rouet}, and {\name Stora} \refI\
carried out the renormalization of the model under investigation without
fermion based on the use of the scheme introduced by {Bogoliubov, Parasiuk,
Hepp}, and Zimmermann (BPHZ) \refIV\ which is not gauge invariant. By this
they succeeded for the first time in giving that model an interpretation in
the framework of a \FoRa\ operator theory by means of the finite mass
renormalization of both the physical and the unphysical degrees of freedom.
Such an interpretation is very convenient for a discussion of asymptotic
problems, e.g.\ the unitarity of the $S$ matrix. This way they have
shown the independence of physical observables on parameters of the gauge
fixing sector. The condition for the application of this concept which can
easily be generalized to other (e.g. non-\Abel) gauge models is the
invariance under a generalized gauge transformation, the {\it BRS
trans\-for\-ma\-tion.\/} This invariance is accomplished by the \SlId\ in
higher orders in perturbation, which governs the renormalization procedure.

Putting emphasis to the consistency of the model under study and facing the
proofs of the unitarity and the fact that physical observables do not
depend on parameters of the gauge fixing sector
this work is therefore based on these later achievements as summarized in
refs.\ \refVI.

Another difference between ref.\ \refGJ\ and this paper is the following.
The spontaneous symmetry breaking makes one degree of freedom of the
{Higgs} field unphysical and massless due to the {Goldstone} theorem. In
order to give it a mass the {'t~Hooft} gauge is used. This choice makes the
introduction of \FaPo\ ghosts necessary because the {Goldstone} boson is
not free.

The subsequent paper is organized as follows. In section 1 the most general
invariant and power-count\-ing renormalizable \LaFu\ is constructed giving the
field content and the BRS transformations. $\sCP$ invariance is imposed which
plays an analogous role as $\sC$ invariance in the \HiKi\ model \refI. Section
2 is devoted to the \SlId\ and the anomaly cancellation. However, it is not
carried through how to fulfill the \SlId\ because of lack of space. This
standard procedure is reviewed for more general cases in refs.\ \refVI\ or
\refII. The consequences of introducing the \WeZu\ counterterms are illustrated
in a certain example for a physical scattering amplitude. Section 3 presents a
closer look at the \Nonre.
\nextkap
\kusout{Tree Approximation}
\aus{The Fields and their BRS Transformations}
The field content of the model is given by a {Dirac} particle $\ps$, a complex
scalar Higgs field $\Hi$ which breaks the symmetry spontaneously, a gauge
field $A_\my$, and the \Geist\ fields $c$ and $\cq$. An auxiliary ({Lagrange}
multiplier) field $b$ is useful to obtain the full BRS invariance off-shell.
The fields have the following power-count\-ing dimensions, graduations, and
ghost numbers
\gl{felder}{\vcenter{\hbox{\vrule\vbox{\tabskip=1pc
                                       \baselineskip=15pt
        \halign{\hfill $#$\hfill &\hfill $#$\hfill &\hfill $#$\hfill
                &\hfill $#$\hfill\cr
                \noalign{\hrule\medskip}
                &{\rm dim}&\vert\>\>\vert&\fapo\cr
                \noalign{\medskip\hrule\medskip}
                \ps ,\;\psq&3/2&1&0\cr
                \Hi&1&0&0\cr
                A^\my&1&0&0\cr
                c&1&1&1\cr
                \cq&1&1&-1\cr
                b&2&0&0\cr
                \noalign{\medskip\hrule\medskip}
                \s&1&1&1\cr
                \noalign{\medskip\hrule}
               }}\vrule}}}

The BRS transformations are
\glalign{brstrafos}
          {     \s\ps_R &= iq_R\,c\ps_R,\ncr{\gln.i}
                \s\ps_L &= iq_L\,c\ps_L, \ncr{ii}
                \s\Hi   &= iq\,c\Hi,\ncr{iii}
                \s A_\my &= -\,\pd_\my c,\ncr{iv}
                \s c     &= 0,\ncr{v}
                \s\cq    &= i\,b,\ncr{vi}
                \s b     &= 0,\ncr{vii}}
where $q_R$, $q_L$, and $q$ are charge numbers.
\aus{Discrete Symmetries}
The anomaly reads
\gl{anomalie}{{\cal A}\propto c\,\ep^{\my\ny\rh\si}\,\pd_\my A_\ny \pd_\rh
                 A_\si,}
thus being {\it even\/} under charge conjugation $\sC$ (where the
\Geist\ $c$ is arbitrarily assumed to be even). Now, the simultaneous charge
conjugation invariance of the action and the anomaly implies a contradiction
due to the renormalized action principle. Hence charge conjugation invariance
must be relinquished if one wants to investigate an anomaly. Therefore the two
chirality parts of the {Dirac} field (eq.\ (\brstrafos .i and ii)) should
transform differently whereby the gauge field couples to the axial fermion
charge, too.

But one can insist upon invariance under time reversal $\sT$ (or $\sCP$
respectively). This discrete symmetry plays a role analogous to $\sC$ in
the \HiKi\ model. Therefore the result is the same non-fermionic part of
the \LaFu\ as in ref.\ \refI\fusn{except for an additional BRS invariant in
ref.\ \refI\ which is excluded here by introduction of the auxiliary field
$\sst b$ in the BRS transformations}.

The phase of the complex field $\Hi$ is chosen such as to give the real part
the quantum numbers of the vacuum which is assumed to be time reversal
invariant. Consequently, the imaginary part is $\sT$ odd. Under BRS
transformations the two parts of the {Higgs} field transform as
\gl{hisp}{\eqalign{\Hi &= {1\over\sqrt{2}}\,\bigl((v + \hi) + i\,\go\bigr) ;\cr
                 \s\hi &= -q\,c\go ,\cr
                 \s\go &=  q\,c\,(v + \hi)\cr}}
with the vacuum expectation value ${1\over\sqrt{2}}v \equiv \vev{\Hi}$.

The following \Herm\ \Lore\ tensors that can be constructed from the fields are
even ($+$) or odd ($-$) under $\sCP$ and $\sT$; space-time arguments and
indices transform respectively.
\gl{disksy}{\vcenter{\hbox{\vrule\vbox{\tabskip=1pc
                                       \baselineskip=15pt
        \halign{\hfill $#$\hfill &\hfill $#$\hfill &\hfill $#$\hfill\cr
                \noalign{\hrule\medskip}
                &\sCP &\sT\cr
                \noalign{\medskip\hrule\medskip}
                \hi ,\;\psq\ps ,\; i\,\psq\pdrlsl\ps ,\;
                i\,\psq\pdrlsl\ga_5\ps ,\; g^{\my\ny}
                &+&+\cr
                \go ,\; i\,\psq\ga_5\ps ,\;\psq\si_{\my\ny}\ps,\;
                \ep_{\my\ny\rh\si},\;
                \pd_\my A_\ny\>{\rm etc.},\; b
                &-&-\cr
                \pd_\my\hi\>{\rm etc.}
                &+&-\cr
                \pd^\my\go\>{\rm etc.},\;\psq\ga^\my\ps ,\;
                \psq\ga^\my\ga_5\ps ,\; A^\my
                &-&+\cr
                \noalign{\medskip\hrule\medskip}
                \s
                &-&-\cr
                c,\; \cq
                &+&+\cr
                \noalign{\medskip\hrule}
               }}\vrule}}}
\aus{\LaFu}
In the framework of BPHZ renormalization \refIV\ the definition of a quantized
model is based on \Zimm's effective \LaFu\
\gl{laeff}{\La_\sub{eff}\mdef\sum_{i=1}^n\la_i{\cal M}_i,}
where ${\cal M}_i$ are \Herm, \Lore\ invariant, homogeneous polynomials of the
model fields, called in the following ``monomials'' by abuse of language. The
parameters $\la_i$ have a perturbative expansion
\gl{lambdasII}{\la_i=\sum_{l=0}^{\infty}\la_i^{(l)}.}
To lowest order they {\it are\/} the renormalized parameters of the model and
they are given by the classical \LaFu. The parameters of higher order ($l>0$)
correspond to local, {\it finite\/} counterterms which compensate for the loop
corrections at the points where the corresponding vertex functions are
renormalized.

The task is to define all parameters $\la_i^{(l)}$ by imposing BRS invariance
and the normalization conditions. In a power-count\-ing renormalizable theory
the number $n$ of the monomials is fixed, finite, and defined by imposing the
power-count\-ing dimensions of the monomials not to exceed the space-time
dimension. The BPHZ regularization besides respects discrete symmetries. Thus
that set of normalization conditions which defines the model in the tree
approximation is necessary and sufficient to define all orders. The most
general ansatz for eq.\ (\laeff) is the one that has the right quantum numbers
and works at the classical level.

A \LaFu\ $\La$
always consists of a {\it gauge invariant\/} physical
part $\La_\sub{inv}$ which depends neither on the ghosts nor on the auxiliary
field, and a gauge fixing and ghost part $i\,\s\Q$ which is unphysical \refVI\
\refIII:
\gl{brandt}{\La = \La_\sub{inv}(\ph) + i\,\s\Q(\ph ,\; c,\; \cq ,\; b).}
The latter part is separately BRS invariant because the BRS operator is
nilpotent
\gl{nilpot}{{\sl s}^2 = 0}
where $\s$ is \Herm.

In addition to the four gauge invariants of the \HiKi\ model there are three
invariants containing the {Dirac} field in this model: two kinetic terms,
$i\,\psq\Drlsl\ps$ and $i\,\psq\Drlsl\ga_5\ps$, and a {Yukawa} coupling,
$\psq_R\ps_L\Hi^* + \psq_L\ps_R\Hi$. The covariant derivative $D$ is given by
\gl{kovabl}{
    \eqalign{
        D_\my\ps_{R\atop L} &\mdef (\pd_\my + iq_{R\atop L}\,
                        A_\my)\,\ps_{R\atop L},\cr
        D_\my\Hi &\mdef (\pd_\my + iq\, A_\my)\,\Hi , \cr}}
where $q=q_L-q_R$ because of the Yukawa coupling.

The non-van\-ish\-ing of the normalization of the gauge field dependent part of
$\Q$ is necessary in order to implement a gauge fixing and to make the
propagation of the gauge field and the Goldstone field well-de\-fin\-ed.
Attention is drawn to the interesting fact that one is led to this in a natural
way by the generality of the BRS invariant ansatz for $\La$. The function $\Q$
is odd under $\sCP$, has dimension $3$, ghost number $-1$, and therefore the
following form
\gl{qgestalt}{\Q=\cq\,\left(-{\al\over 2}\,b+\Ska+\ze\,\go+\thv\,\hi\go\right)}
where $\al$, $\ze$, and $\thv$ are additional free parameters. Thus there are
four unphysical BRS invariants. One may now choose a special class of gauge
fixing without changing the physical content. First of all, one is only
interested in a linear gauge function, i.e.\fusn{This is identical with the
gauge function from ref.\ \refI.}
\gl{vtheta}{\thv = 0.}
Secondly, the mixing propagator $\vevt{\Ska\,\go}_0$ {}\fusn{Here and in the
following $\sst\vevt{\cdot\cdot}_0$ denotes the free two-point function.}
vanishes by choice of
\gl{zzeta}{\ze = \al v.}
This is the so-called ``restricted {'t~Hooft} gauge''. The $b$-propagator is
skillfully ``diagonalized'' by the substitution
\gl{bstrich}{b'\mdef \al\, b-\Ska-\ze\,\go.}

After rescaling $A_\my\rightarrow e\,A_\my$, the \LaFu\ gets the form
\gl{lafuI}{
    \eqalign{
   \La =\>&\halb\,\pd_\my\hi\pd^\my\hi+\halb\,\pd_\my\go\pd^\my\go
            -e\,A^\my\hi\pdrl_\my\go \cr
          &-\viertel\ka{(\hi^2+\go^2)}^2-\ka v\,\hi\,(\hi^2+\go^2)
            -\ka v^2\,\hi^2 \cr
          &+\psq\left({\textstyle {i\over 2}}\,\pdrlsl
            -e\,\Asl\,{1+\ga_5\over 2}
                -m\left(1+{\hi + i\ga_5\,\go\over v}\right)\right)\ps \cr
          &-\viertel F^2
            +\halb e^2v^2\,A\cdot A
                      \left(1+2{\hi\over v}+{\hi^2+\go^2\over v^2}\right) \cr
          &+{\la\over 2e^2}b^{\prime 2}-{\la\over 2}(\Ska)^2
            -{1\over 2}{e^2v^2\over\la}\go^2
                \qquad -ev\,\pd\cdot(A\go) \cr
          &+i\,\cq\left(\d'al
                +{e^2v^2\over\la}\left(1+{\hi\over v}\right)\right)c \cr
   }}
with $\la\mdef e^2\mathbin{/}\al$. The choice of the charge numbers is
$q_R=1,\; q_L=0$.

At this stage one may impose a necessary and sufficient number of
normalization conditions in view of higher orders of perturbation theory,
namely in order to fix the wave functions, the vacuum expectation value of the
physical {Higgs} field ($\vev{\hi}=0$), the free parameters $v$, $\ka$, $m$,
$e$, $\la$, $\ze$, $\thv$, and the ratio of the charge numbers in such a way,
that at the classical level one gets the above \LaFu.

The resulting propagator of the gauge field
\gl{gaugeprop}{\intd{x}e^{i\,p\cdotp x}\,\vevt{A_\my(0)\,A_\ny(x)}_0
               =i\,{-g_{\my\ny}+p_\my p_\ny/e^2v^2\over p^2-e^2v^2+\ips}
               -i\,{p_\my p_\ny/e^2v^2\over p^2-e^2v^2/\la +\ips}}
behaves like $p^{-2}$ for large momentum $p$---it ``scales'' at high energy
and the (non-) renormalizability is read off the vertices alone. In turn
there is a propagating scalar ghost which should however be cancelled by the
Goldstone boson with the propagator
\gl{goprop}{\intd{x}e^{i\,p\cdotp x}\,\vevt{\go(0)\,\go(x)}_0
            ={i\over p^2-e^2v^2/\la+\ips}}
because of BRS invariance. This is quite similar to section~III stage~ii in
ref.\ \refGJ, which would be equivalent to $\ze = 0$ here and also results in a
scaling propagator. The advantage of the restricted {'t~Hooft} gauge is that
the scalar part of the gauge field and the Goldstone boson simultaneously
become massive and do not cause infrared singularities. Thereby one is on the
safe side, for the Goldstone boson ($\go$)~- ghost ($\Ska$) cancellation is not
yet achieved in the non-in\-var\-i\-ant regularization procedure. It is
established by renormalization by means of the \SlId. Besides \FaPo\ fields
have to be introduced for the compensation of the effect of the Goldstone mass.
\nextkap
\kusout{Perturbation Theory of Higher Orders}
\aus{\SlId}
Now the task is to define all parameters $\la_i^{(l)}$ with $l>0$. For that a
generalization of the BRS transformation to higher orders in $\h$ is needed.
Because it is non-lin\-e\-ar in the quantized and interacting fields (i.e.\
composite operators) due to the non-triv\-i\-al \FaPo\ sector one has to add
sources to the \LaFu\ which generate the field transformations:
\gl{laext}{\eqalign{
        \La_\sub{new}&=\La_\sub{old}+q_\al\,\s\ph^\al,\cr
        q_\al\,\s\ph^\al
           &=\si c\go-\ta c(v+\hi)+i\,\rhq c\ps_R-i\,c\psq_R\rh
           -{1\over e}\yp^\my\pd_\my c.\cr}}
Dimensions and charges of the anti-\Herm\ sources $q_\al$ are given by
\gl{quext}{\vcenter{\hbox{\vrule\vbox{\tabskip=1pc
                                   \baselineskip=15pt
        \halign{\hfill $#$\hfill &\hfill $#$\hfill &\hfill $#$\hfill
                &\hfill $#$\hfill\cr
                \noalign{\hrule\medskip}
                &{\rm dim}&\vert\>\>\vert&\fapo\cr
                \noalign{\medskip\hrule\medskip}
                \si,\;\ta,\;\yp^\my&2&1&-1\cr
                \rhq,\;\rh&3/2&0&-1\cr
                \noalign{\medskip\hrule}
             }}\vrule}}\>}
$\La_\sub{new}$ is BRS invariant if one requires $\s q_\al=0$. One has to
introduce further normalization conditions for the new terms.

The generalized BRS transformation of the quantum vertex functional
$\Ga[\ph;q]$
\gl{slavnop}{
        \Sl(\Ga)=\intd{x}\left( {\de\Ga\over\de\si}\,{\de\Ga\over\de\hi}
                +{\de\Ga\over\de\ta}\,{\de\Ga\over\de\go}
                +{\de\Ga\over\de\rhq_\al}\,{\de\Ga\over\de\ps_\al}
                +{\de\Ga\over\de\rh_\al}\,{\de\Ga\over\de\psq_\al}
                +{\de\Ga\over\de\yp^\my}\,{\de\Ga\over\de A_\my}
                +i\,b{\de\Ga\over\de\cq}\right)}
is the {Slavnov} transformation. The \SlId\ which generalizes the BRS
invariance reads
\gl{slid}{\Sl(\Ga)=0.}
The renormalized action principle for the Slavnov transformation
\gl{swipri}{\Sl(\Ga)=\Sigma^{(l)}+\ord{\h^{l+1}}}
is valid; $\Sigma[\ph;q]$ is a local functional of ghost number one. There is
now a local differential operator $\Bn$ with
\gl{konsbedIII}{\Bn\,\Sigma^{(l)}=0}
which is {\it nilpotent\/} and identical with the classical BRS operator $\s$
if applied to the fields, $\Bn\,\ph^\al=\s\ph^\al$. Eq.\ (\konsbedIII) is the
\WeZu\ consistency condition \refWZ.

Its solution has the form \refIII
\gl{sigestalt}{\Sigma^{(l)}=A+\Bn\,\ga,\qquad A=\intd{x}{\cal A}\ne\Bn\ga',}
where $A$ and $\ga$ are of formal order $\h^l$. This is quite analogous to the
form of the \LaFu\ (\brandt). $A$ is a ``genuine'' anomaly. The terms
$\Bn\,\ga$ are not actually anomalous, i.e.\ in the sense of an inconsistency,
because it is possible to absorb them recursively to all orders $l$ in the
course of the renormalization procedure. In this way the \SlId\ can be
fulfilled if there is no term $A$. The question of the existence of a
``genuine'' anomaly $A$ is a local cohomology problem and just answered by
algebraic methods \refIII. It is a property of the model and does not depend
on the renormalization procedure.
\aus{Cancellation of the Anomaly}
In this model (\anomalie) is no ``genuine'' anomaly---and there is no
``genuine'' anomaly at all because there is a function of the {Higgs} boson
which transforms into $\hbox{const.}\cdot c$, i.e.\ its phase relative to the
vacuum
\gl{omegaI}{\om\mdef v\,\arctan{\go\over v+\hi}.}
It is asymptotically identical with the {Goldstone} field which has the
inhomogeneous BRS transformation. In the case $v=0$ this
expansion does not exist. But with $v>0$ the model contains a field $\om$
which transforms as
\gl{omtrafo}{\s\om=-v\,c,}
so the anomaly is $\Bn$ exact:
\gl{kuerz}{cF\widetilde{F}=\Bn\left(-{\om\over v}F\widetilde{F}\right).}
The BRS invariance violating vertex insertion $\Sigma^{(1)}_\sub{chiral}$
contributed by the diagrams shown in \abbI\ is computed to be
\gl{anocount}{\Sigma^{(1)}_\sub{chiral}
        =\intd{x}{e^2\over 48\pi^2}cF\widetilde{F}
        =\Bn\intd{x}{-e^2\over 48\pi^2v}\om F\widetilde{F}
        =\Bn\,\ga^{(1)}_\sub{chiral}.}
(The upper vertices of the diagrams are terms of the insertion.)
\midinsert
\vskip 7 true cm
\centerline{\eightrm\abbI :\quad The diagrams contributing to the chiral
                                 anomaly.}
\centerline{\eightrm\vrule width 0pt height 7pt
       The triangles of solid lines denote the sum over}
\centerline{\eightrm\vrule width 0pt height 7pt
       the two fermion loops with resp.\ opposite direction.}
\bigskip
\endinsert

The counterterm $-\ga_{\sub{chiral}}$ consists of infinitely many monomials:
\gl{klamone}{{\cal M}_{i'}=\hi^{m_{i'}}\go^{2n_{i'}+1}F\widetilde{F},
        \qquad m_{i'},\;n_{i'}\>\in\{0,1,2,\dots\}}
($n\not<\infty$ any more in eq.\ (\laeff)). They are all power-count\-ing {\it
non\/}-re\-nor\-mal\-i\-za\-ble.
\aus{Discussion}
Before analyzing the renormalizability a physical scattering amplitude
involving the unphysical poles and the anomaly shall be considered in order to
see their cancellation on the one hand and the \hien\ behaviour which is the
origin of the \Nonre\ on the other hand.

\midinsert
\vskip 5.7 true cm
\centerline{\eightrm\abbpaar :\quad The diagrams with the unphysical poles
                                    and the anomaly}
\centerline{\eightrm\vrule width 0pt height 7pt
                                    which contribute to the fermion pair
                                    annihilation.}
\centerline{\eightrm\vrule width 0pt height 7pt
       The dashed line denotes the Goldstone boson propagator}
\centerline{\eightrm\vrule width 0pt height 7pt
       and the full circle is the Wess-Zumino term.}
\bigskip
\endinsert
The corresponding contributions of leading order to the fermion pair
annihilation (\abbpaar) have also been considered by {\name Gross} and {\name
Jackiw} \refGJ. As they already pointed out, the residue of the {\it
unphysical\/} pole (here at $(p+q)^2=e^2v^2/\la$ in contrast to ref.\ \refGJ)
is proportional to
$$i\,{(p+q)}_\la\,\Ga^{(1)}_{A_\la(0)\ft{A}_\ny(p)\ft{A}_\si(q)}
           +ev\,\Ga^{(1)}_{\go(0)\ft{A}_\ny(p)\ft{A}_\si(q)}.$$
This sum vanishes because the \SlId\ in the one-loop approximation (or the Ward
identity resp.) is valid and thus the Goldstone boson ($\go$)~- ghost
($\Ska$) cancellation occurs. Diagrammatically this is so because the
well-known anomalous violation of PCAC is cancelled by the \WeZu\ counterterm
${e^2\over 48\pi^2v}\,\go F\widetilde{F}$ which contributes locally to
$\Ga^{(1)}_{\go\ft{A}_\ny\ft{A}_\si}$ with $-\,{e^2\over
12\pi^2v}\ep^{\my\ny\rh\si}p_\my q_\rh$.

Whereas the \hien\ behaviour of the complete pair annihilation amplitude {\it
without\/} \WeZu\ counterterm is as expected, the counterterm causes trouble
however. In order to see this one can apply the criterion of ``\treun i\-ty''
introduced by {\name Cornwall, Levin,} and {\name Tiktopoulos}
\refCLT\fusn{who derived gauge invariance from \hien\ unitarity bounds}. It
says that any physical $T$ matrix element in the tree approximation of a
\LaFu\ model with $N$ initial and final particles must not grow faster than
$E^{4-N}$ in the limit of high center-of-mo\-men\-tum energy $E$ and fixed,
non-ex\-cep\-tion\-al angles in order not to exceed the Froissart bounds.

Because in the tree approximation the model is power-count\-ing
renormalizable it is formally ``\treun y''. However, the considered amplitude
which belongs to the first order in $\h$ does not fulfill the criterion. In
the \hien\ limit the counterterm contributes a factor proportional to
$\ep^{0\ny\rh\si}{\hat p}_\rh\,E^2$ with ${\hat p}^\my=(0, {\bf p}/\grad{{\bf
p}})^\my$ and $\lim_{E\to\infty}p^\my={E\over 2}\,(1, {\bf p}/\grad{{\bf
p}})^\my$. Therefore the corresponding contribution to the $T$ amplitude grows
like $E^1$ in contrast to all other contributions which tend at most to
constants (apart from logarithmical factors; note that spinors grow like
$\sqrt{E}$).

Thus the requirement of BRS invariance of this anomalous gauge model
violates the Froissart bounds in the one-loop approximation. One may view
this as indication for the fact that perturbation theory fails above some
energy \refLS\ because above that energy the \WeZu\ contribution which is
of first order in $\h$ dominates for instance the pair annihilation.
\nextkap
\kusout{Non-Renormalizability}
As claimed in ref.\ \refCLT\ the on-shell property ``\treun i\-ty'' is a
necessary condition for renormalizability, thus the model cannot be
renormalizable.

An essential input of the BPHZ regularization is the power-count\-ing formula
\gl{powcount}{\eqalign{
                      d(\Gamma)&=4-B-{3\over 2}F-\sum_a(4-\de_a) \cr
\hbox{with}\qquad \de_a&=b_a+{3\over 2}f_a+D_a+c_a \cr}}
($B$, $F$: external boson, fermion line; $b_a$, $f_a$ count the bosons or
fermions resp.\ at the vertex, $D_a$ is the power of momenta belonging to the
vertex, $c_a$ is a non-negative degree of oversubtraction, and the index $a$
numbers the vertices) which specifies the degree of superficial divergence
of a \Feyn\ diagram or sub-diagram $\Gamma$.
Conventionally, the degree $c_a$ of oversubtraction is chosen such as
to equate $\de_a=4$ and to determine the superficial divergence of a whole
vertex function by the external lines only. This is possible in
power-count\-ing renormalizable theories, i.e.\ if all monomials ${\cal
M}_i$ have field dimension $b_a+{3\over 2}f_a+D_a\leq 4$.

In this model one has to consider the
full formula (\powcount). Vertex functions which are convergent up to first
order in $\h$ may consequently become undetermined with growing order of
perturbation theory because they contain the vertices of eq.\ (\klamone)
which make some contributing (sub-)diagrams divergent. One must therefore
add further counterterms of field dimension $>4$ to the effective \LaFu\
with respect to these vertex functions.

The result 
is that already in second order perturbation theory new normalization
conditions become necessary. In order to see this it is sufficient to search
for new BRS invariants of dimension $>4$ which are linear combinations of
(both old and new) counterterms of the corresponding order. In order to find
out new counterterms one looks for diagrams which become divergent just
because of the vertices (\klamone) and which have external fields
corresponding to really new terms. It was found by inspection that the unique
divergent functions of second order whose counterterms are not determined by
invariants of dimension $\leq 4$ have the structure shown in \abbII. The upper
vertices originate from the monomials of (\klamone) and are of formal order
$\h^1$, i.e. just determined by eq.\ (\anocount) without further corrections.
The BRS invariants which have to be renormalized read
\gl{nrcountform}{
        \psq\si^{\my\ny}(v+\hi+i\ga_5\,\go)\ps F_{\my\ny}f(\grad{\Hi}^2).}
Because of the arbitrariness of the function $f$ they are even infinitely
many. The model is therefore \nonre\ from the second order onwards in the
following sense: The first order is still well-defined by the symmetry and the
classical normalization conditions only but the higher orders cannot be defined
by a finite set of normalization conditions at all.
\midinsert
\vskip 6 true cm
\centerline{\eightrm\abbII :\quad Diagrams of second order}
\centerline{\eightrm\vrule width 0pt height 7pt
        which make further normalization conditions necessary}
\bigskip
\endinsert
One may ask if this simple and superficial inspection based on
power-count\-ing in the framework of the BPHZ renormalization scheme is
actually satisfactory since one could attempt to reach renormalizability by
pushing all ``non-re\-nor\-mal\-iza\-bil\-i\-ties'' onto one field by
formulating the model in terms of the Higgs phase $\om$ and the radial part
$\rhv$ (with $\vev{\rhv}=0$) instead of Cartesian variables as fundamental
fields. The kinetical \LaFu\ term for the Higgs field would read
$$ \halb\,\pd_\my\rhv\pd^\my\rhv
  +\halb  \left(1+{\rhv\over v}\right)^2
          \left(\pd_\my\om\pd^\my\om
                -2ev\,A^\my\pd_\my\om
                +e^2v^2\,A^2\right) $$
and the Yukawa coupling
$$-m\,\left(1+{\rhv\over v}\right)\,\psq\,\exp\left(i\ga_5{\om\over v}
                                              \right)\,\ps.$$
Because of loss of power-count\-ing renormalizability the first order already
needs additional normalization conditions in this case, and therefore this
alternative is \nonre\ from the first order onwards. However, the number of
\WeZu\ counter\-terms shrinks to one, cf.\ eqs.\ (\kuerz) and (\anocount).
Consequently there is only a {\it finite\/} number of normalization conditions
which are additionally necessary for each order in $\h$.

One could now try to give $\om$ power-count\-ing dimension $0$ by modifying the
bilinear part of the \LaFu\ with a view of proving the renormalizability of
the complete model together with the \WeZu\ term by power counting. However,
exactly this modification is not possible without destroying the consistency
because it would necessarily mean a propagator for $\om$
\gl{omprop}{\intd{x}e^{i\,p\cdotp x}\,\vevt{\om(0)\,\om(x)}_0
            \propto {i\over p^2-{m_1}^2+\ips}-{i\over p^2-{m_2}^2+\ips}}
which behaves like $p^{-4}$ for large $p$. Thus the model would be afflicted
with an additional ghost field due to the ``wrong'' sign of the one part of
the propagator. This ghost must be physical since an unphysical modification
cannot alter the \hien\ behaviour of a physical scattering amplitude like the
discussed fermion pair annihilation.
\nextkap
\kusout{Conclusion and Discussion}
The \Abel\ \HiKi\ model with a chirally coupled fermion is an anomalous gauge
model. It has been perturbatively quantized preserving the \SlId\ in arbitrary
orders in $\h$. Since there is a \WeZu\ counterterm this is possible because of
the inhomogeneous gauge transformation of the Goldstone boson due to the SSB.
Since BRS invariance is established one can easily transcribe the proof of the
unitarity from ref.\ \refI\ or \refVI\ to this model. Thus a consistent,
unitary, \Loko, four-di\-mension\-al, massive gauge model is obtained.

It has been shown that the model is not renormalizable in the sense that it
cannot be defined by a finite number of normalization conditions. The net
result of imposing BRS invariance is only that one has pushed the \Nonre\
which at first appeared as bad-be\-hav\-ed vector boson propagator to second
order in $\h$. According to this it is expected that perturbation theory
breaks down above a certain energy.

The shown cancellation of the anomaly is suggested as a method of quantizing
anomalous gauge models in the sense of ref.\ \refJ. The role of the
``chiral field'' is played here by the Higgs phase $\om$. In ref.\ \refJ\
the method of decoupling {\it additional\/} anomaly-cancelling fermions is
mentioned. That method has already been explicitly applied to the model
under study in ref.\ \refGJ. The decoupling of the fermion with the other
axial-vector charge by sending the corresponding Yukawa coupling to
infinity leaves behind the same \WeZu\ term as in this work. However, it is
stressed that the decoupling method differs from the demonstrated one by
the fact that the resulting effective action contains other \nonre\ terms
in addition to the \WeZu\ term (see also ref.\ \refDF). Even if one is only
interested in the perturbation series up to the first order which is still
meaningful\fusn{i.e.\ completely defined by the normalization conditions
determining the tree approximation} the two methods give different results.
The decoupling method results in an effective low-en\-er\-gy theory per
definition. Apparently, the shown quantization method which is a {\it
``minimal''\/} anomaly cancellation in the sense that it manages without
additional effective terms has no corresponding interpretation for high
energies.
\bigskip\bigskip
\aus{Acknowledgements}
I wish to thank F.\ Brandt, H.\ D.\ Dahmen, N.\ Dragon, R.\ Kretschmer, S.\
Marculescu, O.\ Piguet, M.\ Reuter, K.\ Sibold, and L.\ Szymanowski for
helpful and illuminating discussions.
\nextkap
\kusout{References}
{\parindent=2pc
\item{\refJ}{\name Jackiw R}: Update on Anomalous Theories, MIT preprint
CTP {\bf 1436} (1986)
\item{\refGJ}{\name Gross D J, Jackiw R}: Phys.\ Rev.\ {\bf D6} (1972) 477
\item{\refLA}{\name Lee B W}: Phys.\ Rev.\ {\bf D5} (1972) 823
\item{}{\name Adler S L}: Lectures on Elementary Particles and Quantum
Field Theory (Eds.\ Deser S, Grisaru M, Pendelton H; MIT Press 1970)
\item{\refWZ}{\name Wess J, Zumino B}: Phys.\ Lett.\ {\bf 49B} (1974) 52
\item{\refI}{\name Becchi C, Rouet A, Stora R}: Comm.\ math.\ Phys.\ {\bf 42}
(1975) 127
\item{\refVI}{\name Piguet O, Rouet A}: Phys.\ Rep.\ {\bf 76C} (1981) 1
\item{}{\name Becchi C}: Relativity, Groups and Topology II (Les Houches
1983 Session XL, Eds.\ DeWitt B S, Stora R; North-Holland 1984) 789
\item{\refII}{\name Sibold K}: Lecture Notes in Physics {\bf 303} (Eds.\
Breitenlohner P, Maison D, Sibold K; 1987) 32
\item{\refIV}{\name Lowenstein J H}: Renormalization Theory (Eds.\
Velo G, Wightman A S; 1975) 94
\item{}{\name Lowenstein J H}: Seminars on Renormalization Theory (given in
Maryland 1972)
\item{\refIII}{\name Brandt F, Dragon N, Kreuzer M}: Nucl.\ Phys.\ {\bf B332}
(1990) 224
\item{\refCLT}{\name Cornwall J M, Levin D N, Tiktopoulos G}: Phys.\ Rev.\
{\bf D10} (1974) 1145
\item{\refLS}{\name Llewellyn Smith C H}: Phys.\ Lett.\ {\bf 46B} (1973)
233
\item{\refDF}{\name D'Hoker E, Farhi E}: Nucl.\ Phys.\ {\bf B248} (1984) 59

}
\vfill
\eject
\end